\documentclass{aa}  

\usepackage{graphicx}
\usepackage{txfonts}
\usepackage[colorlinks=true,citecolor=blue,linkcolor=blue,urlcolor=blue]{hyperref}
\usepackage{longtable}
\usepackage{multirow}
\usepackage{upgreek}
\usepackage{orcidlink}  
\def\ha{H$\upalpha$}
\def\hb{H$\upbeta$}
\def\pa{Pa$\upalpha$}
\def\jwst{JWST}
\def\hst{HST}

\begin{document} 

   \title{A first look at a complete view of spatially resolved star formation at $1<z<1.8$ with JWST NGDEEP+FRESCO slitless spectroscopy}

    \titlerunning{Spatially resolved Paschen-alpha vs. H-alpha}

    \authorrunning{Matharu et al.}


   \author{Jasleen Matharu
           \inst{1}\fnmsep
           \inst{2}\fnmsep
           \inst{3}\thanks{{Corresponding author; \email{jamatharu@mpia.de}}}\orcidlink{0000-0002-7547-3385},
           Lu Shen
           \inst{4}\fnmsep
           \inst{5},
           Irene Shivaei
           \inst{6},
           Pascal A. Oesch
           \inst{1}\fnmsep
           \inst{2}\fnmsep
           \inst{7},
           Casey Papovich
           \inst{4}\fnmsep
           \inst{5},
           Gabriel Brammer
           \inst{1}\fnmsep
           \inst{2}\orcidlink{0000-0003-2680-005X}, 
           Naveen A. Reddy\inst{8}\orcidlink{0000-0001-9687-4973},
           Yingjie Cheng
           \inst{9}\fnmsep
           \inst{10}\orcidlink{0000-0001-8551-071X},
           Pieter van Dokkum\inst{11}\orcidlink{0000-0002-8282-9888},
           Steven L. Finkelstein
           \inst{12}\fnmsep
           \inst{13}\orcidlink{0000-0001-8519-1130},
           Nimish P. Hathi\inst{14}\orcidlink{0000-0001-6145-5090},
           Jeyhan S. Kartaltepe\inst{15}\orcidlink{0000-0001-9187-3605},
           Anton M. Koekemoer\inst{14}\orcidlink{0000-0002-6610-2048},
           Jorryt Matthee\inst{16}\orcidlink{0000-0003-2871-127X},
           Nor Pirzkal\inst{14}\orcidlink{0000-0003-3382-5941},
           Stephen M.~Wilkins\inst{17}\fnmsep
           \inst{18}\orcidlink{0000-0003-3903-6935},
           Michael A. Wozniak\inst{8}\orcidlink{0000-0002-1033-3656}, and
           Mengyuan Xiao\inst{7}\orcidlink{0000-0003-1207-5344}}

   \institute{{Cosmic Dawn Center, Copenhagen, Denmark}
        \and
        {Niels Bohr Insitute, University of Copenhagen, Jagtvej 128, 2200 Copenhagen, Denmark}
        \and
        {Max-Planck-Institut f\"ur Astronomie, K\"onigstuhl 17, D-69117 Heidelberg, Germany}
        \and
        {Department of Physics and Astronomy, Texas A\&M University, College Station, TX, 77843-4242, USA}
        \and
        {George P.\ and Cynthia Woods Mitchell Institute for Fundamental Physics and Astronomy, Texas A\&M University, College Station, TX, 77845-4242, USA}
        \and
        {Centro de Astrobiolog\'{i}a (CAB), CSIC-INTA, Carretera de Ajalvir km 4, Torrej\'{o}n de Ardoz, 28850, Madrid, Spain}
        \and
        {Department of Astronomy, University of Geneva, Chemin Pegasi 51, 1290 Versoix, Switzerland}
        \and
        {Department of Physics and Astronomy, University of California,
Riverside, 900 University Avenue, Riverside, CA 92521, USA}
        \and
        {Department of Astronomy, University of Washington, Seattle, WA 98195, USA}
        \and
        {University of Massachusetts Amherst, 710 North Pleasant Street, Amherst, MA 01003-9305, USA}
        \and
        {Astronomy Department, Yale University, 52 Hillhouse Ave,
New Haven, CT 06511, USA}
        \and
        {Department of Astronomy, The University of Texas at Austin, Austin, TX, USA}
        \and
        {Cosmic Frontier Center, The University of Texas at Austin, Austin, TX 78712, USA}
        \and
        {Space Telescope Science Institute, 3700 San Martin Drive, Baltimore, MD 21218, USA}
        \and
        {Laboratory for Multiwavelength Astrophysics, School of Physics and Astronomy, Rochester Institute of Technology, 84 Lomb Memorial Drive, Rochester, NY 14623, USA}
        \and
        {Institute of Science and Technology Austria (ISTA), Am Campus 1, A-3400 Klosterneuburg, Austria}
        \and
        {Astronomy Centre, University of Sussex, Falmer, Brighton BN1 9QH, UK}
        \and
        {Institute of Space Sciences and Astronomy, University of Malta, Msida MSD 2080, Malta}}

   \date{Received XXX; accepted XXX}

 
  \abstract 
   {The previously inaccessible star formation tracer Paschen-Alpha ({\pa}) can now be spatially resolved by JWST NIRCam slitless spectroscopy in distant galaxies up to cosmic noon. In the first study of its kind, we combine JWST NGDEEP NIRISS and FRESCO NIRCam slitless spectroscopy to provide the first direct comparison of spatially resolved dust-obscured (traced by {\pa}) versus unobscured (traced by {\ha}) star formation across the main sequence. We stack {\pa} and {\ha} emission-line maps,  along with stellar continuum images at both wavelengths of 31 galaxies at $1<z<1.8$ in three bins of stellar mass spanning $7.7\leqslant\mathrm{log}(M_{*}/\mathrm{M}_{\odot})<11$. Surface brightness profiles are measured and equivalent width (EW) profiles computed. Increasing {\pa} and {\ha} EW profiles with galactocentric radius across all stellar masses probed provide direct evidence for the inside-out growth of galaxies both via dust-obscured and unobscured star formation for the first time. For galaxies predominantly on the main sequence ($\mathrm{log}(M_{*}/\mathrm{M}_{\odot})\geqslant8.8$), a weakly positive ($0.1\pm0.1$) {\pa}/{\ha} line profile gradient as a function of radius is found at $8.8\leqslant\mathrm{log}(M_{*}/\mathrm{M}_{\odot})<9.9$ with a negative ($-0.4\pm0.1$) {\pa}/{\ha} line profile gradient as a function of radius found at the highest stellar masses of $9.9\leqslant\mathrm{log}(M_{*}/\mathrm{M}_{\odot})<11.0$. Low mass galaxies ($7.7\leqslant\mathrm{log}(M_{*}/\mathrm{M}_{\odot})<8.8$) with predominantly high specific star formation rates (sSFRs) relative to the main sequence are also found to have a negative ($-0.5\pm0.1$) {\pa}/{\ha} line profile gradient as a function of radius.
   Our results demonstrate that while inside-out growth via star formation is ubiquitous across the main sequence just after cosmic noon, centrally concentrated dust attenuation is not. Along with other recent work in the literature, our findings now motivate future studies of resolved SFR and dust attenuation profiles in large samples of individual cosmic noon galaxies across the main sequence, to understand the intrinsic scatter in spatially resolved star formation.}


   \maketitle
%
\section{Introduction}

Understanding how galaxies build up their stellar mass over cosmic time is a central goal in observational astrophysics, providing critical insight into how galaxies develop their structure. Tracking ongoing star formation in galaxies is therefore key to revealing the physical processes that regulate galaxy growth. Spatially resolving star formation in galaxies is the only way to determine {\it where} star formation is occurring in galaxies, thereby revealing how galaxies assemble their stellar mass and evolve.

Multiple observational tracers spanning from the X-ray to radio regime of the electromagnetic spectrum can be used to trace ongoing star formation in galaxies over various timescales. Of these, emission-line tracers provide a near-instantaneous measure of star formation due to their capability of tracing emission from the ionised gas surrounding massive young stars (\citealt{Kennicutt2012} and references therein). The {\ha} emission line has been a popular choice for observing star formation in both nearby and distant galaxies due to its easy access in the rest-frame optical and high intrinsic brightness. Spatially resolved {\ha} emission-line maps have revealed that from the end of reionisation ($z\sim5.3$) to $z\sim0.5$, galaxies grow ``inside-out'' by forming new stars at larger and larger galactocentric radii over time \citep{Nelson2012, Nelson2016, Wilman2020, Matharu2022, Matharu2024, Shen2024, Danhaive2025} but may be experiencing centralised bursty star formation into the epoch of reionisation \citep{Stephenson2025}.

Despite the conveniences of using {\ha} as a star formation tracer, its susceptibility to dust attenuation leads to a systematic underestimation of star formation rates (e.g. \citealt{Whitaker2014, Shivaei2016}). In cases of low to moderate dust attenuation, {\ha}-derived star formation rates (SFRs) can be corrected using the Balmer decrement, whereby the {\ha}/{\hb} flux ratio together with an assumed dust attenuation curve can be used to quantify the amount of dust attenuation towards star-forming regions \citep{Calzetti1999, Osterbrock2006}. Integrated Balmer decrements increase with stellar mass and SFR \citep{Wild2011,Dominguez2012,Momcheva2013,Price2014,Reddy2015,Shivaei2015,Pirzkal2024,Sandles2024} but remain constant with redshift: a trend that is difficult to physically explain 
\citep{Shapley2022,Shapley2023}. The few high redshift ($z\geqslant1$) studies that exist on spatially resolved Balmer decrements of typical star-forming galaxies find no shape evolution in the dust attenuation profiles with redshift at fixed stellar mass to help explain this trend, with predominantly centrally concentrated negative profiles found \citep{Nelson2016a,Matharu2023}.

A gold standard alternative for obtaining a complete view of star formation in distant galaxies lies with the Paschen-Alpha ({\pa}) emission line. By virtue of being in the rest-frame near infra-red (NIR), it is insensitive to dust attenuation. {\pa} emission is therefore capable of tracing star formation in more heavily dust-obscured regions that would be optically thick to Balmer lines \citep{Kennicutt2012}. Telluric OH emission, thermal emission, and its intrinsic faintness have made this line inaccessible for ground-based observatories. The Spitzer Space Telesecope could access {\pa} in high redshift galaxies, but only in rare cases of strong gravitational lensing \citep{Papovich2009, Shipley2016}. Its rest-frame wavelength (1.875$\mu\mathrm{m}$) lies beyond the WFC3/IR wavelength regime covered by the Hubble Space Telescope (HST), making it inaccessible from space as well. Now, with the launch of {\it} JWST, the {\pa} emission line has become accessible in multiple observational modes out to cosmic noon (NIRSpec slit spectroscopy: \citealt{Reddy2023}, NIRCam medium-band imaging: \citealt{Williams2023,Suess2024} and NIRCam slitless spectroscopy: \citealt{Liu2024,Neufeld2024}). So far there is evidence to suggest that {\pa}-derived SFRs could be $\sim25\%$ higher than those derived from Balmer lines \citep{Reddy2023} with SFR({\pa}/{\ha}) increasing with stellar mass \citep{Liu2024}. At $1<z<1.6$, the {\pa} versus stellar continuum effective radius is positive and increases with stellar mass, providing the first direct evidence for the inside-out growth of galaxies via dust-obscured star formation \citep{Liu2024}.

In this paper, we provide the first direct spatially resolved comparison of unobscured versus dust-obscured star formation using {\ha} and {\pa} emission-line maps for the same galaxies at $1<z<1.8$ residing in the Hubble Ultra Deep Field (HUDF) using {\jwst} NIRISS and NIRCam slitless spectroscopy from the Next Generation Deep Extragalactic Exploratory Public Survey (NGDEEP, \citealt{Bagley2024}) and the First Reionization Epoch Spectroscopically Complete Observations (FRESCO, \citealt{Oesch2023}). Our sample is described in Section~\ref{sec:sample} and our unique methodology for enabling direct comparisons between NIRCam and NIRISS slitless spectroscopy is detailed in Section~\ref{sec:methodology}. We outline our results in Section~\ref{sec:results} and discuss their physical interpretations in Section~\ref{sec:discussion}. A summary of our findings is given in Section~\ref{sec:summary}.

All magnitudes quoted are in the AB system, logarithms are in base 10, and we assume a $\Lambda$CDM cosmology with $\Omega_{m}=0.307$, $\Omega_{\Lambda}=0.693$, and $H_{0}=67.7$~kms$^{-1}$~Mpc$^{-1}$ \citep{Planck2015}.


\section{Sample}
\label{sec:sample}

   \begin{figure}
   \centering
   \includegraphics[width=\columnwidth]{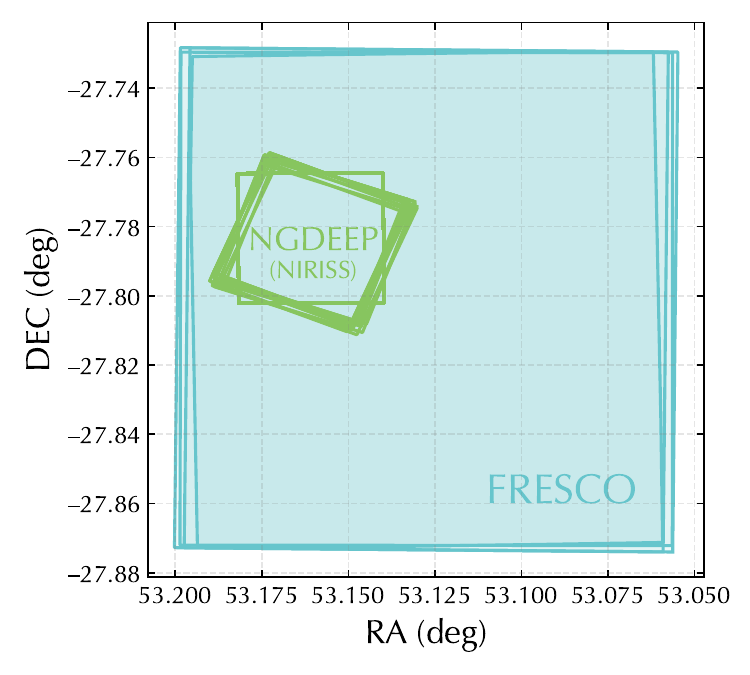}
      \caption{The FRESCO and NGDEEP NIRISS observing footprints on the sky.}
              
         \label{fig:footprint}
   \end{figure}

We describe the data we use in this section (Section~\ref{sec:data}), how the {\ha} and {\pa} emission-line maps are generated (Section~\ref{sec:maps}) and how the final sample is selected (Section~\ref{sec:sample_selection}).

\subsection{Data and data processing}
\label{sec:data}

We use data from the {\jwst} Cycle 1 programs NGDEEP and FRESCO. NGDEEP is a large treasury survey with 122.6 hours of NIRISS F115W, F150W, and F200W imaging and slitless spectroscopy observations centered on the HUDF and 96.4 hours of parallel NIRCam imaging of the HUDF-Par2 field: the deepest {\hst} ACS F814W field on the sky. FRESCO is a medium survey with 53.8 hours of F444W NIRCam imaging and slitless spectroscopy observations of the GOODS-S and GOODS-N CANDELS fields. In this work, we use the NIRISS data from NGDEEP and the GOODS-S NIRCam data from FRESCO, that overlaps over the HUDF (Figure~\ref{fig:footprint}). For more details on the NGDEEP and FRESCO surveys, we refer the reader to \cite{Bagley2024} and \cite{Oesch2023}, respectively.

The Grism Redshift and Line Analysis Software (\texttt{grizli}; \citealt{Grizli2022}) was used to process the imaging and slitless spectroscopy together. \texttt{grizli} downloads and pre-processes the raw data from the Mikulski Archive for Space Telescopes (MAST). A basis set of template flexible stellar population synthesis models (\texttt{FSPS}; \citealt{Conroy2009,Conroy2010}) is projected to the pixel grid of the 2D grism exposures using the spatial morphology from the imaging. Observed spectra are then fit with 2D template spectra with non-negative least squares. To break redshift degeneracies, a line complex template is used and the grism redshift is taken to be where the $\chi^{2}$ is minimised across the grid of trial redshifts input by the user.

\subsection{Emission-line maps}
\label{sec:maps}
The $\sim2-4$ times smaller bandwidths of the NIRISS filters and the order of magnitude lower resolving power of the NIRISS grisms compared to F444W NIRCam mean that the NGDEEP NIRISS slitless spectra are significantly shorter than the FRESCO NIRCam slitless spectra. The standard \texttt{grizli} approach of forward-modelling the direct imaging to construct a contamination model for each 2D grism exposure is therefore computationally feasible for NIRISS grism observations and is used to process the NGDEEP NIRISS observations (\citealt{Shen2024} and references therein). In the \texttt{grizli}-processing for the FRESCO observations, grism spectra are instead median-filtered to remove the continuum (\citealt{Matharu2024} and references therein). Emission-line maps are created by inputting the wavelength of the line in the \texttt{grizli}-extraction process, which creates continuum-subtracted narrow-band maps given a specific wavelength.

\subsection{Sample selection}
\label{sec:sample_selection}

The FRESCO grism spectra and \texttt{grizli} data products are visually inspected to verify {\pa} detections by the \texttt{grizli} software, as follows. 
Sources with {\pa} signal-to-noise ratio, \mbox{SNR~$>3$} are initially selected from the \texttt{grizli} catalog for visual inspection. Two inspectors independently assess each using the \texttt{SpecVizitor} software\footnote{\href{https://github.com/ivkram/specvizitor}{https://github.com/ivkram/specvizitor}}, classifying them into five categories based on line fit quality and redshift probability distribution: (1) contaminated, (2) no lines, (3) uncertain line fit quality, (4) robust single-line fit with consistent F444W morphology, and (5) robust multiple line fits. Sources flagged as robust (quality 4 or 5) by both reviewers are labeled "robust"; those flagged by only one are "semi-robust". A weighting scheme accounts for inspector bias where necessary. For this study, we select robust ($\sim30\%$) and semi-robust ($\sim 40\%$) sources, further limited to those with {\pa} \mbox{SNR~$>5$}, yielding a final catalog of 508 robust {\pa} detections in GOODS-S. These detections are cross-matched with the 1153 {\ha} detections in the NGDEEP \texttt{grizli} extraction, yielding 43 matches. The NGDEEP grism spectra and \texttt{grizli} data products for this sample are visually inspected alongside the FRESCO \texttt{grizli} data products to verify the {\ha} detections and the quality of both the {\ha} and {\pa} emission-line maps. 35 galaxies are identified with good quality {\ha} and {\pa} emission-line maps. For more details on the quality criteria for emission-line maps, we refer the reader to Section 2.3 of \cite{Matharu2024}.

\subsubsection{Stellar masses and star formation rates}
\label{sec:masses_sfms}

   \begin{figure}
   \centering
   \includegraphics[width=\columnwidth]{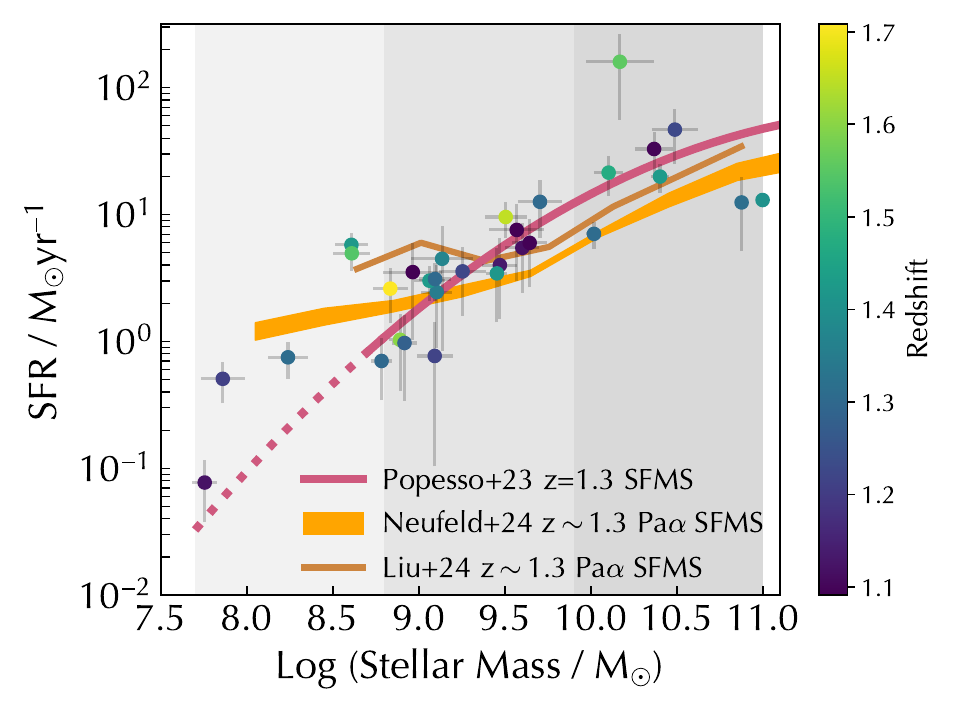}
      \caption{Star formation main sequence of our sample measured using SED fitting (see Section~\ref{sec:masses_sfms}). The shaded grey regions delineate our stellar mass bins for the stacking analysis. SFRs and the \cite{Popesso2023} SFMS include dust corrections. The dotted-line indicates extrapolation. The brown line and orange region delineate the FRESCO {\pa} SFMS from \cite{Liu2024} and \cite{Neufeld2024} respectively.}
              
         \label{fig:SFMS}
   \end{figure}

The Spectral Energy Distribution (SED) fitting code CIGALE \citep{Boquien2019,Yang2020} is used to derive stellar masses and SFRs for our galaxies in the NGDEEP NIRISS field. CANDELS \citep{Koekemoer2011,Grogin2011} and JADES DR2 \citep{Eisenstein2023, Rieke2023, Williams2023} photometry spanning 0.43-4.8 $\mu$m is used with the grism redshifts determined from the \texttt{grizli} NGDEEP extraction. The \cite{Calzetti1999} and \cite{Cardelli1989} dust laws with $R_{V}=3.1$ are used for attenuating the stellar continuum and emission-lines, respectively. Three Active Galactic Nuclei (AGNs) are identified and removed from the sample by cross-matching to \cite{Lyu2022} with a 1 arcsecond radius. AGN activity can lead to {\ha} and {\pa} emission. Because we are only interested in studying star formation traced by {\ha} and {\pa}, we remove AGN from our sample that would otherwise complicate the physical interpretation of our results. We then remove one galaxy from our sample for which we are unable to derive a stellar mass due to unavailable photometry. This leaves us with 31 galaxies that constitute our final sample. For full details on the SED fitting procedure -- including a comparison of {\ha} versus SED based SFRs -- we refer the reader to \cite{Shen2025}. The star formation main sequence (SFMS) of our final sample is shown in Figure~\ref{fig:SFMS} along with the three stellar mass bins (grey shaded regions) we divide our sample into for forthcoming analysis. The majority of the sample lies along the literature SFMS from \cite{Popesso2023} for the median redshift of our sample, suggesting we are primarily studying typical main sequence galaxies at this redshift.

\section{Methodology}
\label{sec:methodology}

In this section we describe how we prepare our data to enable the first direct comparison of {\ha} and {\pa} spatial profiles in the same galaxies at $1<z<1.8$.

\subsection{Point-spread function matching}
\label{sec:PSFs}

   \begin{figure}
   \centering
   \includegraphics[width=\columnwidth]{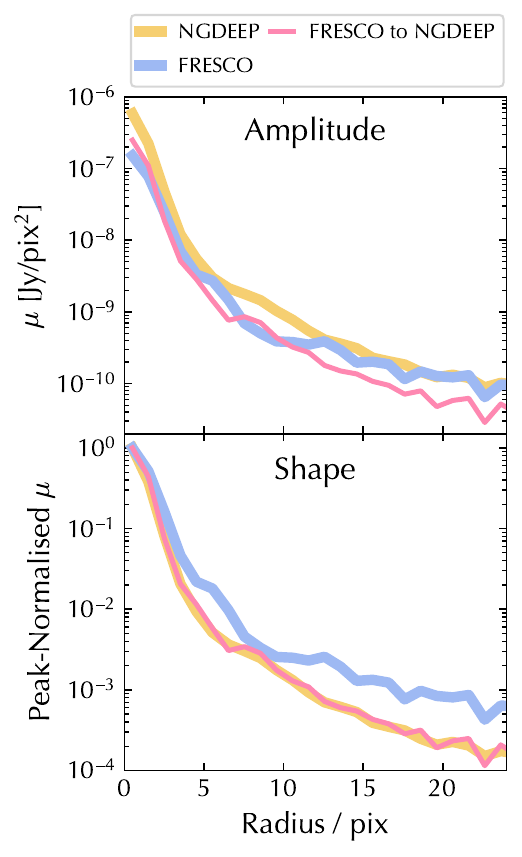}
      \caption{PSF surface brightness profiles using the same star in the F150W NGDEEP NIRISS and F444W FRESCO NIRCam imaging. The thin pink line shows the PSF profile after PSF-matching (see Section~\ref{sec:PSFs}) the FRESCO PSF to the NGDEEP PSF. The amplitude (top panel) of the PSF is maintained and its shape (bottom panel) is appropriately altered.}
              
         \label{fig:PSFs}
   \end{figure}

The Point Spread Function (PSF) of an instrument describes how light spreads on a detector for a point source and so defines the resolution limit of the detector. The {\ha} and {\pa} emission-line maps used in our study are obtained from slitless spectroscopy taken with two different instruments on-board JWST, which have different PSFs. Since the primary science goal of our study requires us to provide a direct comparison of {\ha} and {\pa} spatial profiles in the same galaxies at these redshifts for the first time, the difference in PSFs needs to be accounted for. The PSF of a detector can be measured using observations of a star. The gold and blue thick lines in Figure~\ref{fig:PSFs} show the surface brightness profiles of the same star in the NGDEEP and FRESCO \texttt{grizli} direct image thumbnails which both have the same pixel scale and orientation. Not only do the amplitudes differ (top panel), but so do the shapes, more explicitly shown in the bottom panel of Figure~\ref{fig:PSFs}, where the profiles have been peak-normalised. 

To address this problem, we PSF-match the higher resolution FRESCO observations to the lower resolution NGDEEP observations. This is done by supplying \texttt{pypher} \citep{Boucaud2016} with the direct image thumbnails of the two stars in the same units to generate a matching kernel between the FRESCO and NGDEEP observations. The \texttt{astropy} package \citep{TheAstropyCollaboration2018} is then used to convolve the FRESCO F444W imaging and {\pa} maps with their respective matching kernels. The pink lines in Figure~\ref{fig:PSFs} show the result of this convolution on the FRESCO PSF profile. The amplitude of the FRESCO PSF is retained (top panel) while the shape is appropriately altered to match that of the NGDEEP PSF (bottom panel).

\subsection{Stacking}
\label{sec:stacking}

   \begin{figure}
   \centering
   \includegraphics[width=\columnwidth]{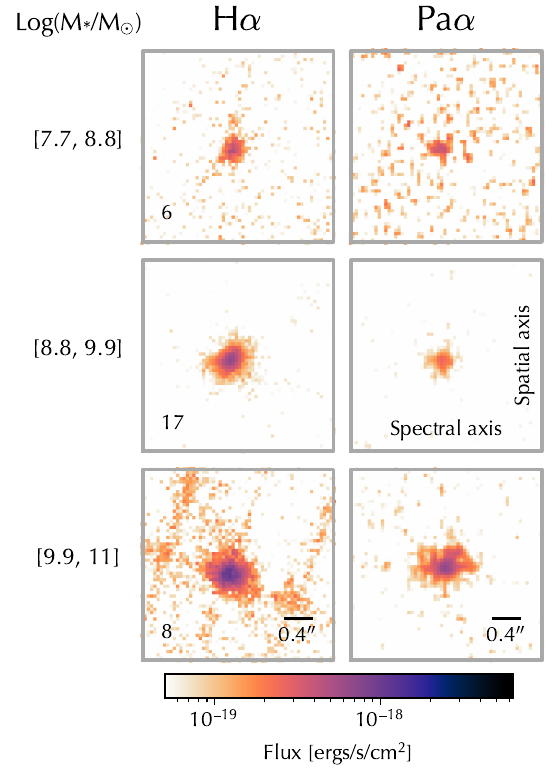}
      \caption{{\ha} (first column) and {\pa}-to-NGDEEP convolved (second column) stacks for our sample in three stellar mass bins (rows). The number of galaxies is shown in the bottom left corner of each {\ha} stack thumbnail. Each thumbnail is $60\times60$ pixels with 1 pixel = 0.05$^{\prime\prime}$. Emission-line maps are generated such that the spectral axis is along the $x$-direction, while the $y$-direction is the spatial axis (see Section~\ref{sec:profiles}).}
              
         \label{fig:stacks}
   \end{figure}

To boost the signal-to-noise ratio as a function of galactocentric radius, we stack the {\ha} and {\pa} emission-line maps along with the direct imaging in three bins of stellar mass for our sample: {$7.7\leqslant\mathrm{Log}(M_{*}/\mathrm{M}_{\odot})<8.8$}, {$8.8\leqslant\mathrm{Log}(M_{*}/\mathrm{M}_{\odot})<9.9$} and {$9.9\leqslant\mathrm{Log}(M_{*}/\mathrm{M}_{\odot})<11.0$}. Whilst {\ha} is predominantly detected within the F150W filter wavelength range for our sample, there are a few cases where it is detected in the F200W wavelength range. We therefore use the NGDEEP F150W and F200W \texttt{grizli}-generated thumbnails of the star used for the PSF matching (Section~\ref{sec:PSFs}) between the two datasets to construct PSF stacks for our NGDEEP continuum and {\ha} stacks. Matching kernels are then created between the FRESCO PSF in F444W to these PSF stacks that are used to PSF-match (see Section~\ref{sec:PSFs}) the F444W continuum and {\pa} stacks to the NGDEEP stacks. Full details of our stacking process can be found in \cite{Matharu2022}. In summary, neighbouring sources are masked in both the direct image and emission-line map thumbnails using the \texttt{grizli}-generated segmentation map thumbnails for each galaxy. Pixels are weighted using the \texttt{grizli}-generated inverse variance map thumbnails and each galaxy is weighted by its total flux in the corresponding direct image filter such that no single bright galaxy dominates the stack. Direct image and emission-line map thumbnails are summed and exposure-corrected using the sum of their inverse variance maps. Each pixel is equal to $0.05$ arcseconds and the dimensions of all thumbnails and the final stacks are $160\times160$~pixels. Zoomed-in regions of our final {\ha} and PSF-matched {\pa} stacks can be seen in Figure~\ref{fig:stacks}.

\subsection{Surface brightness profiles}
\label{sec:profiles}

Due to the high spectral resolution of the NIRCam grism, maximal spatial information from the {\pa} emission-line maps can only be obtained along the cross-dispersion axis, hereafter the ``spatial axis'' (see Figure~\ref{fig:stacks} and \citealt{Matharu2024} for more details). The FRESCO observations were taken at a position angle (PA) of zero (see Figure~\ref{fig:footprint}), leading to the spatial axis being along the vertical axis of all \texttt{grizli}-processed imaging, grism spectroscopy and emission-line maps. The NGDEEP \texttt{grizli} data processing was run with a rotation such that all final data products have PA=0. Therefore, the spatial axis of the FRESCO emission-line maps and imaging directly correspond to 
the vertical axis of the NGDEEP emission-line maps and imaging. We therefore measure surface brightness profiles along the central vertical strip of all our stacks with \texttt{MAGPIE}\footnote{https://github.com/knaidoo29/magpie/}. These profiles form the basis of our analysis which we present in Section~\ref{sec:results}.

\section{Results}
\label{sec:results}

   \begin{figure*}
   \centering
   \includegraphics[width=\textwidth]{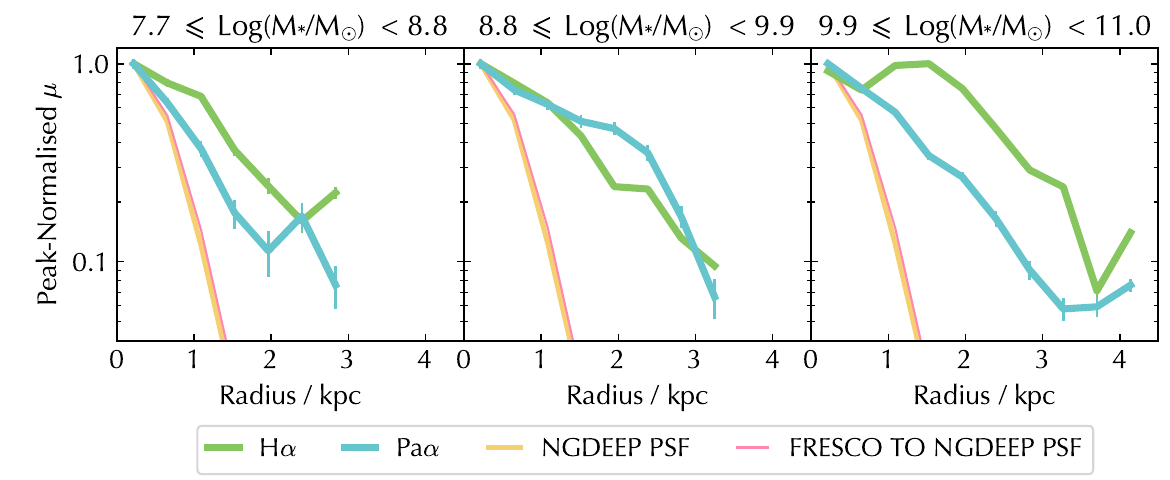}
      \caption{Peak-normalised {\ha} (green) and {\pa} (turquoise) surface brightness profiles for our stacks shown in Figure~\ref{fig:stacks}. Both profiles are more extended than their PSFs, demonstrating they are well-resolved. At the lowest and highest masses, the {\ha} emission is more extended than the {\pa} emission, whereas their spatial profiles are similar for our middle stellar mass bin, at least out to $\sim2$ kiloparsecs after which the {\pa} emission is more extended than the {\ha} emission.}
         \label{fig:profiles_norm}
   \end{figure*}

Figure~\ref{fig:profiles_norm} shows the peak-normalised surface brightness profiles for the {\ha}, {\pa} and PSF stacks along the central vertical strip of the spatial axis. The emission-line map profiles are always more extended than their PSF profiles, demonstrating they are all spatially resolved by JWST NIRCam and NIRISS. For our lowest and highest mass bins, the {\ha} emission is more extended than the {\pa} emission. Whereas for our middle mass bin, {\ha} and {\pa} have similar spatial profiles out to $\sim2$~kpc after which {\pa} emission is more extended than the {\ha} emission out to 3~kpc.

   \begin{figure*}
   \centering
   \includegraphics[width=\textwidth]{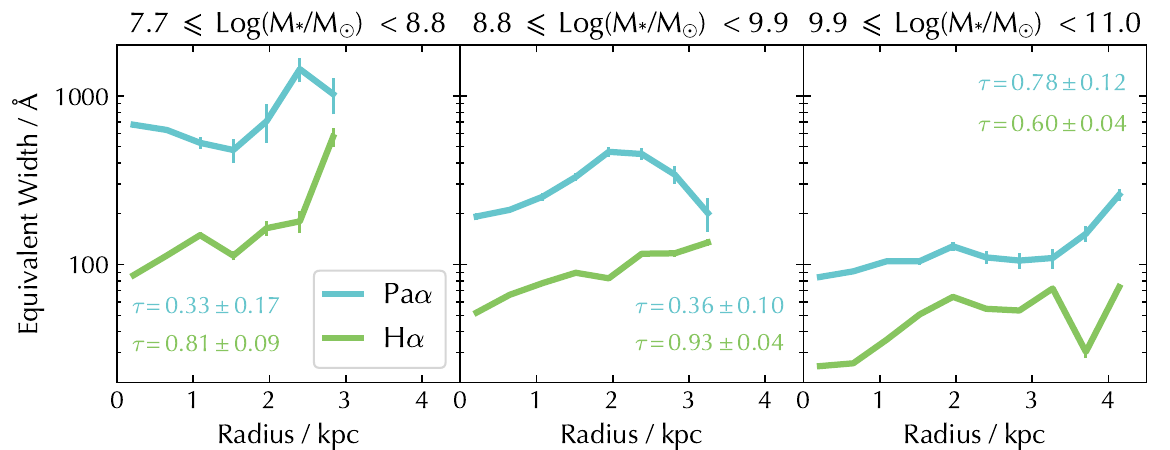}
      \caption{{\pa} (turquoise) and {\ha} (green) equivalent width (EW) profiles for our stacks with their Kendall's Tau ($\tau$) correlation statistic. All EW profiles display a positive radial gradient, demonstrating star-forming galaxies at these redshifts and stellar masses are growing inside-out {\it both} via dust-obscured ({\pa}) and unobscured ({\ha}) star formation.}
              
         \label{fig:profiles}
   \end{figure*}

   \begin{figure*}
   \centering
   \includegraphics[width=\textwidth]{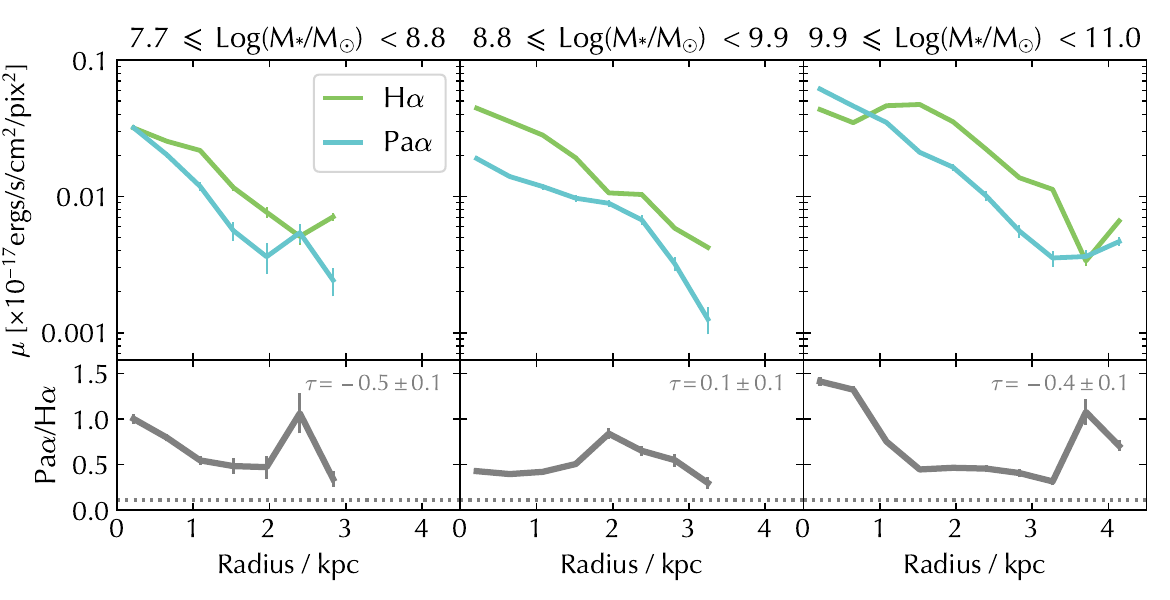}
      \caption{Surface brightness profiles for our {\ha} (green) and {\pa} (turquoise) stacks (top row) with their ratios (bottom row) and Kendall's Tau ($\tau$) correlation statistic. The horizontal dotted line shows the expected {\pa}/{\ha} line ratio for Case B recombination, $T_{e}=10,000$~K and $n_{e}=100~\mathrm{cm}^{-3}$ based on photoionisation modelling with \texttt{CLOUDY} \citep{Ferland2017}. Declining {\pa}/{\ha} with galactocentric radius in our highest and lowest mass bins highlights the additional, dust-obscured star formation {\pa} is able to trace closer to galactic centers that decreases towards the outskirts of these galaxies.}
              
         \label{fig:linefluxes}
   \end{figure*}

We show the quotient of the emission-line map profiles with their respective continuum profiles in Figure~\ref{fig:profiles}. Known as the equivalent width (EW), this quantity provides an absolute measurement for the strength of the emission line. We use the Kendall's Tau ($\tau$) statistic to measure the correlation between equivalent width and radius. Errors on $\tau$ are computed using Monte Carlo sampling by drawing 1000 random samples of each line profile and their errors from a normal distribution with a standard deviation equal to the measurement errors. The error on $\tau$ is then the standard deviation of the 1000 $\tau$ measurements obtained. Both EW({\ha}) and EW({\pa}) have positive profile gradients as a function of radius for all stellar mass bins. EW({\pa}) always has a higher normalisation than EW({\ha}).

Figure~\ref{fig:linefluxes} shows the emission line flux profiles for our {\ha} and {\pa} stacks and their ratios. As demonstrated by the $\tau$ statistic, our {\pa}/{\ha} profiles have negative gradients with increasing radius for the lowest and highest stellar mass bins, but a mildly positive gradient with increasing radius for our middle stellar mass bin.

In the next section, we will discuss the physical interpretation of our results.

\section{Discussion}
\label{sec:discussion}
The most massive and short-lived ($\sim10$~Myr) O-type stars emit the strongest ultraviolet (UV) radiation, capable of stripping hydrogen atoms of their electrons. The recombination of these electrons with ionised hydrogen atoms further from the centre of these stars leads to the emission of hydrogen recombination lines such as {\ha} in the rest-frame optical and {\pa} in the rest-frame near-infrared (NIR). Whilst the {\ha} emission line is intrinsically stronger than the {\pa} emission line -- and therefore more than eight times brighter (Case B recombination, $T_{e}=10^{4}$~K) -- it is more susceptible to dust attenuation. Therefore, {\ha} is often regarded as a good tracer of unobscured ongoing star formation only. {\pa} is dubbed a ``gold standard'' tracer of ongoing star formation due to its capability in tracing more heavily dust-obscured ongoing star formation \citep{Kennicutt2012}.

\subsection{Galaxy growth}
\label{sec:galaxy_growth}
The equivalent width (EW) profiles of each of these emission lines (Figure~\ref{fig:profiles}) measure their strength with respect to their underlying continuum as a function of galactocentric radius. The normalisation of both EW profiles fall with increasing stellar mass, consistent with a fall in sSFRs (\citealt{Popesso2023} and references therein) and the increasing contribution of older stellar populations to the stellar continuum both in the optical and NIR. In general, all EW profiles are positive for all three stellar mass bins, providing direct evidence for the inside-out growth of galaxies via star formation after cosmic noon, confirmed here for the first time with both {\ha} and {\pa} (see also \citealt{Liu2024} for {\pa} versus continuum size analysis).

Areas that are often missed in EW({\ha}) profiles are optically thick regions at the rest-frame wavelength of {\ha} (6565{\AA}) and/or regions that have higher nebular dust attenuation which reduce the {\ha} line flux more relative to the stellar attenuation reducing the continuum flux. In the first case, both the {\ha} line flux and continuum is affected. The second case could be the case for star-forming regions (e.g. \citealt{Calzetti1999,ForsterSchreiber2009,Yoshikawa2010,Mancini2011,Wuyts2011,Wuyts2013, Kashino2013, Kreckel2013,Price2014,Reddy2015,Bassett2017, Theios2019,Koyama2019a,Greener2020, Wilman2020,Rodriguez-Munoz2021}) and leads to an underestimation in EW({\ha}). In the next section, we therefore discuss a cleaner approach to comparing unobscured versus obscured star formation using our measurements.

\subsection{Unobscured vs. obscured star formation}
It is impossible to gain deeper insight into spatially resolved unobscured versus obscured star formation by directly comparing {\ha} and {\pa} EW profiles. This is because of complications introduced by differential dust attenuation of the {\ha} line and the continuum at 6565{\AA} (see Section~\ref{sec:galaxy_growth}) as well as differing relative contributions of different stellar populations to the continuum in the rest-frame optical versus the NIR. We therefore instead directly compare the {\ha} and {\pa} line profiles with their ratios in Figure~\ref{fig:linefluxes}. On average, line flux increases with stellar mass inline with increasing SFRs along the main sequence (see Figure~\ref{fig:SFMS}). In the lowest and highest mass bins, the {\pa}/{\ha} profiles have declining trends with radius from the center towards the expected value of {\pa}/{\ha} in the absence of dust attenuation (horizontal dotted line, Case B recombination, $T_{e}=10,000$~K and $n_{e}=100~\mathrm{cm}^{-3}$ based on photoionisation modelling with \texttt{CLOUDY} \citep{Ferland2017}). In our lowest mass bin, we are predominantly probing galaxies with high sSFRs and in our highest mass bin, those with the highest SFRs in our sample (see Figure~\ref{fig:SFMS}). This trend in the {\pa}/{\ha} profiles suggests that the dust in the central regions of these galaxies is preferentially attenuating the {\ha} flux relative to the {\pa} flux. Our result at the highest stellar masses is consistent with those of many ALMA studies (see discussion in \citealt{Shen2023}) and recent results found using JWST MIRI imaging by \cite{Shen2023} and \cite{Magnelli2023}. The rate of galaxy growth via star formation with radius measured using {\ha} emission alone in the central regions ($\sim$1.5~kpc) of such galaxies could therefore be overestimated.

Contrastingly, the relatively flat {\pa}/{\ha} line profile of the main sequence galaxies in our middle mass bin suggests that not all main sequence galaxies at cosmic noon that are growing inside-out via star formation have centrally concentrated dust attenuation. This more complicated history of star formation for the central and outer regions of galaxies is supported by the recent results of \cite{Shen2024} who examined spatially resolved EW({\ha}), sSFR and age for 19 main sequence galaxies at $0.6<z<2.2$ of comparable stellar masses ({$9\leqslant\mathrm{Log}(M_{*}/\mathrm{M}_{\odot})<11$}) to our sample in the same (NGDEEP NIRISS) field (see also consistent results in \citealt{Matharu2023, Liu2023, Liu2024, Maheson2025}). Whilst the majority (84\%) of their galaxies have positive EW({\ha}) profiles supporting the inside-out growth scenario, the sSFR and star formation history (SFH) profiles suggest at least one rapid star formation episode is responsible for forming the bulge, but smoothly varying SFHs for the disk suggest longer timescale inside-out growth. Our results along with those of \cite{Shen2024} put forth a more complex picture for spatially resolved star formation at and around cosmic noon. Large sample resolved studies of individual galaxies at cosmic noon across the main sequence are now required to understand the intrinsic scatter in how star formation progresses in galaxies.

\section{Summary}
\label{sec:summary}
By processing overlapping JWST NGDEEP NIRISS and FRESCO NIRCam slitless spectroscopy in the HUDF, we have provided the first direct comparison of spatially resolved unobscured (traced by {\ha}) and dust-obscured (traced by {\pa}) ongoing star formation in 31 galaxies at $1<z<1.8$ (Section~\ref{sec:results}). Our main conclusions are listed below.

\begin{enumerate}
    \item Positive EW({\pa}) and EW({\ha}) profiles demonstrate that main sequence galaxies just after cosmic noon are growing inside-out {\it both} via dust-obscured and unobscured star formation.

    \item At the highest stellar masses and sSFRs, {\pa}/{\ha} falls with increasing galactocentric radius, likely tracing decreasing dust-obscured star formation relative to unobscured star formation towards larger radii.

    \item The relatively flat {\pa}/{\ha} line profile for our stack of galaxies with stellar masses $8.8\leqslant\mathrm{log}(M_{*}/\mathrm{M}_{\odot})<9.9$ suggests centrally concentrated dust attenuation is not ubiquitous across the main sequence just after cosmic noon in galaxies growing inside-out via star formation.

\end{enumerate}

Our work puts forth a proof-of-concept for using NIRISS and NIRCam slitless spectroscopy together to accomplish more detailed resolved studies of high redshift galaxies. Together with other similar works in the literature, this work motivates a strong case for detailed, resolved studies of large samples of individual galaxies across the main sequence at cosmic noon to understand the intrinsic scatter in spatially resolved star formation.

\begin{acknowledgements}
      JM is grateful to the Max Planck Society for the MPIA Prize Fellowship and to the Cosmic Dawn Center for the DAWN Fellowship. This work is based on observations made with the NASA/ESA/CSA James Webb Space Telescope. The raw data were obtained from the Mikulski Archive for Space Telescopes at the Space Telescope Science Institute, which is operated by the Association of Universities for Research in Astronomy, Inc., under NASA contract NAS 5-03127 for \textit{JWST}. These observations are associated with JWST Cycle 1 GO programs \#1895 and \#2079. Support for both programs was provided by NASA through a grant from the Space Telescope Science Institute, which is operated by the Associations of Universities for Research in Astronomy, Incorporated, under NASA contract NAS5-26555. 
      The Cosmic Dawn Center (DAWN) is funded by the Danish National Research Foundation under grant DNRF140.
      This work has received funding from the Swiss State Secretariat for Education, Research and Innovation (SERI) under contract number MB22.00072, as well as from the Swiss National Science Foundation (SNSF) through project grant 200020\_207349.
      This work benefited from support from the George P. and Cynthia Woods Mitchell Institute for Fundamental Physics and Astronomy at Texas A\&M University. CP thanks Marsha and Ralph Schilling for generous support of this research.

\end{acknowledgements}


\bibliography{library_grizli2}{}
\bibliographystyle{aa}

\end{document}